\documentclass[twocolumn,epjc3,final]{svjour3}  
\smartqed  
\RequirePackage{graphicx}
\RequirePackage{mathptmx}      
\RequirePackage{latexsym,amssymb,amsmath}
\RequirePackage[colorlinks,citecolor=blue,urlcolor=blue,linkcolor=blue]{hyperref}
\journalname{Eur. Phys. J. C}

\def\prl{Physical Review Letters}
\def\prd{Physical Review D}
\def\aap{Astron. Astroph.}
\def\apj{Astroph. J. }
\def\apjl{Astroph. J. }
\def\mnras{Mon. Not. R. Astron. Soc.}

\begin{document}

\title{Kerr black hole energy extraction, irreducible mass feedback, and the effect of captured particles charge}

\author{J. A. Rueda\thanksref{e1,addr1,addr2,addr3,addr4,addr5} \and R. Ruffini\thanksref{e2,addr1,addr2,addr6} 
}

\thankstext{e1}{e-mail: jorge.rueda@icra.it}
\thankstext{e2}{e-mail: ruffini@icra.it}

\institute{ICRANet, Piazza della Repubblica 10, I--65122 Pescara, Italy \label{addr1}
           \and
           ICRA, Dipartimento di Fisica, Sapienza Universit\`a di Roma, P.le Aldo Moro 5, I--00185 Rome, Italy \label{addr2}
           \and
           ICRANet-Ferrara, Dipartimento di Fisica e Scienze della Terra, Universit\`a degli Studi di Ferrara, Via Saragat 1, I--44122 Ferrara, Italy \label{addr3}
           \and
           Dipartimento di Fisica e Scienze della Terra, Universit\`a degli Studi di Ferrara, Via Saragat 1, I--44122 Ferrara, Italy \label{addr4}
           \and
           INAF, Istituto de Astrofisica e Planetologia Spaziali, Via Fosso del Cavaliere 100, 00133 Rome, Italy \label{addr5}
           \and
           INAF, Viale del Parco Mellini 84, 00136 Rome  Italy\label{addr6}
}

\date{Received: date / Accepted: date}

\maketitle

\begin{abstract}
We analyze the extraction of the rotational energy of a Kerr black hole (BH) endowed with a test charge and surrounded by an external test magnetic field and ionized low-density matter. For a magnetic field parallel to the BH spin, electrons move outward(inward) and protons inward(outward) in a region around the BH poles(equator). For zero charge, the polar region comprises spherical polar angles $-60^\circ\lesssim \theta \lesssim 60^\circ$ and the equatorial region $60^\circ\lesssim \theta \lesssim 120^\circ$. The polar region shrinks for positive charge, and the equatorial region enlarges. For an isotropic particle density, we argue the BH could experience a \textit{cyclic} behavior: starting from a zero charge, it accretes more polar protons than equatorial electrons, gaining net positive charge, energy and angular momentum. Then, the shrinking(enlarging) of the polar(equatorial) region makes it accrete more equatorial electrons than polar protons, gaining net negative charge, energy, and angular momentum. In this phase, the BH rotational energy is extracted. The extraction process continues until the new enlargement of the polar region reverses the situation, and the cycle repeats. We show that this electrodynamical process produces a relatively limited increase of the BH irreducible mass compared to gravitational mechanisms like the Penrose process, hence being a more efficient and promising mechanism for extracting the BH rotational energy.  
\keywords{gamma-ray bursts: general \and BH physics}
\end{abstract}

\maketitle


\section{Introduction} \label{sec:1}

This paper discusses a specific electrodynamic process to extract the Kerr black hole (BH) rotational energy and analyze its efficiency. For this task, taking into account and differentiating the concepts of extractable energy, extracted energy, and the inevitable increase of the BH irreducible mass in processes interacting with the BH are essential. This fact has been recently exemplified for the gravitational Penrose process \cite{1969NCimR...1..252P} in \cite{2024arXiv240508229R} (hereafter Paper I) and in \cite{2024arXiv240510459R} (hereafter Paper II). Paper I assesses the efficiency of the single Penrose process, while Paper II analyzes the case of repetitive processes. Paper II conclusion is dramatic: applying the Penrose processes repetitively either poorly reduces the BH spin, hence the BH rotational energy, leaving most of it yet to be extracted, or it can approach a final Schwarzschild BH state, but by converting the BH rotational energy into irreducible mass. In either case, very little or even no energy extraction occurs. The former case results from Penrose processes occurring near the BH horizon, and the latter near the ergosphere border. Papers I and II have revealed that the nonlinear increase of the BH irreducible mass is responsible for the Penrose process's inefficiency.

The main consequence of the above result is that any Kerr BH rotational energy extraction process must account for the irreducible mass increase effect. Therefore, the efficiency of any BH energy extraction process is ultimately linked to its ability to approach reversibility, i.e., to cause an as-low-as-possible increase of the BH irreducible mass.

In the context of the Wald electromagnetic field \cite{1974PhRvD..10.1680W} for a Kerr BH immersed in an asymptotically aligned (with the BH spin), uniform test magnetic field, it was shown in Ref. \cite{2023EPJC...83..960R} that the capture and ejection of positively and negatively charged particles by a Kerr BH can extract its energy when the BH captures more particles with negative than positive energy (and angular momentum). These conditions are attainable in an ionized anisotropic medium with density increasing toward the equator. The process showed the BH extracted energy increases more than the irreducible mass, representing a first step towards an efficient process of BH energy extraction. {Indeed, in the sequence of gravitational Penrose processes of Paper II, the increase of the BH irreducible mass relative to the decrease of the mass becomes as large as $\sim 500\%$, which contrasts with the $\sim 1$--$10\%$ relative change obtained in the electrodynamical process in \cite{2023EPJC...83..960R}.}

This paper extends the above analysis by accounting for the fact that the BH also gains charge. As a first step in understanding the process, we consider the BH charge to be a test charge, i.e., it affects the exterior electromagnetic field but not the spacetime geometry, which will still be given by the Kerr metric. The main new result is that including the BH (test) charge allows the BH energy extraction for isotropic surrounding particle density and is more efficient than the previous case. 

The paper is organized as follows. Section \ref{sec:2} discusses the concepts of \textit{extractable} energy, \textit{extracted} energy, and the role of the BH irreducible mass in differentiating the two. Section \ref{sec:3} describes the electromagnetic field structure leading to outgoing and ingoing charged particles. In Section \ref{sec:4}, we calculate the energy and angular momentum of the particles captured by the BH. Section \ref{sec:5} estimates the total change in the BH parameters mass, angular momentum, and irreducible mass. Finally, we outline the conclusions, consequences, and future research directions from the paper's results in Section \ref{sec:6}.

\section{Extractable, extracted energy, and irreducible mass} \label{sec:2}

Let us start by recalling the concept of \textit{extractable energy} (see, e.g., \cite{2021A&A...649A..75M,2022ApJ...929...56R})
\begin{equation}\label{eq:Eext}
    E_{\rm ext} \equiv M - M_{\rm irr},
\end{equation}
where $M$ and $M_{\rm irr}$ are the BH mass and irreducible mass. Unless otherwise specified, we use geometric ($G=c=1$) units throughout. The relation between the BH mass, angular momentum ($J$), charge ($Q$), and irreducible mass is dictated by the Christodoulou-Ruffini-Hawking BH mass-energy formula
\cite{1970PhRvL..25.1596C,1971PhRvD...4.3552C,1971PhRvL..26.1344H}
\begin{equation}\label{eq:Mbh}
M^2 = \left(M_{\rm irr} + \frac{Q^2}{4 M_{\rm irr}^2}\right)^2 + \frac{J^2}{4 M^2_{\rm irr}}.
\end{equation}
The BH horizon is $r_+ = M + \sqrt{M^2 - a^2-Q^2}$, being $a = J/M$ the BH angular momentum per unit mass, so $M_{\rm irr}$ can be readily written as
\begin{equation}\label{eq:Mirr}
    M_{\rm irr} = \frac{1}{2}\sqrt{2 M r_+ -Q^2}.
\end{equation}
For a Schwarzschild BH ($Q=0$ and $J=0$), $M_{\rm irr} = M$, so $E_{\rm ext}=0$. For the extreme Kerr BH ($Q=0$ and $J=M^2$), $M_{\rm irr} = M/\sqrt{2} \approx 0.71 M$, leading to $E_{\rm ext} = (1-1/\sqrt{2})M \approx 0.29 M$. For an extreme Reissner-Nordstr\"om BH ($J=0$ and $Q=M$), $M_{\rm irr} = M/2$, so $E_{\rm ext} = 0.5 M$. Thus, a non-rotating BH has no energy to be extracted, while up to $29\%$ ($50\%$) of the mass of an extremely rotating (charged) BH could be extracted. However, astrophysical processes have to deal with the BH surface area increase theorem  \cite{1971PhRvL..26.1344H}
\begin{equation}\label{eq:S}
dS_+\geq 0, \quad S_+= 4\pi (r_+^2 + a^2) = 16\pi M_{\rm irr}^2,
\end{equation}
which implies that $d M_{\rm irr}^2 \geq 0$ for any process acting on the BH. Hence, the extractable energy {given by Eq. (\ref{eq:Eext}) is the maximum amount of energy that can be \textit{extracted}. If an energy extraction process reduces the BH mass by an amount $dM$, then the extracted energy is defined as
\begin{equation}\label{eq:Eextracted}
     dE_{\rm extracted} \equiv -dM,
\end{equation}
and from Eq. (\ref{eq:Eext}), the extractable energy changes by 
\begin{equation}\label{eq:dEext}
    dE_{\rm ext} = dM - dM_{\rm irr} = -dE_{\rm extracted}- dM_{\rm irr}.
\end{equation}
Thus, the extracted energy can approach the maximum possible value, the extractable energy, only if the process of extraction occurs without increasing the BH irreducible mass, namely if it is reversible in the Christodoulou-Ruffini sense \cite{1970PhRvL..25.1596C,1971PhRvD...4.3552C}, i.e., if $dM_{\rm irr} = 0$.}

Reversibility is hard to approach, so in general, we seek \textit{efficient} processes able to extract the BH energy causing a relatively small change of the irreducible mass \cite{2023EPJC...83..960R,2023ARep...67S..93R}, i.e.,
{
\begin{equation}\label{eq:dMirrcond}
    |dM| = dE_{\rm extracted} \gg dM_{\rm irr},
\end{equation}
such that $dE_{\rm extracted} = -dM \approx dE_{\rm ext}$.}

Let us now focus on the case of a Kerr BH. Equation (\ref{eq:Mirr}) tells that an infinitesimal change in the BH mass ($dM$) and angular momentum ($dJ$) leads to a change in the horizon surface area
\begin{equation}\label{eq:dMirr}
    dS_+ = 16\pi\,d M_{\rm irr}^2= 32\pi M_{\rm irr}^2 \frac{d M - \Omega_+ dJ}{\sqrt{M^2-a^2}},
\end{equation}
where $\Omega_+ = a/(2 M r_+)$. The condition (\ref{eq:dMirrcond}) is not trivial. It challenges energy extraction processes by particles or fields as they could convert the BH rotational energy into irreducible mass rather than in energy extracted. 

When the BH captures a particle of energy $E$ and angular momentum $L$, from Eq. (\ref{eq:dMirr}) we have $d M_{\rm irr}^2 \propto E-\Omega_+L \propto |p^r|_+$, being the latter the radial momentum of the particle crossing the horizon \cite{2018PhRvD..98l3002L,2021PhRvD.104h4059G,2023EPJC...83..960R,2023ARep...67S..93R}. Therefore, a reversible transformation is achieved only by capturing particles \textit{grazing} the horizon \cite{1970PhRvL..25.1596C,1971PhRvD...4.3552C}. Otherwise, $|p^r|_+ >0$, $dS_+>0$ and $dM_{\rm irr}^2 >0$, so $M_{\rm irr}$ necessarily increases.

We have already recalled the results of Paper II, which shows the high irreversibility of a repetitive Penrose process, even approaching the limit of being completely irreversible when it occurs near the ergosphere, i.e., approaching a complete conversion of the BH rotational energy into the irreducible mass without energy extracted: $dE_{\rm ext} \to - dM_{\rm irr}$ and $dE_{\rm extracted} = -dM \to 0$.

We now turn to the electrodynamical processes. We recall that the electrostatic potential for charged particles allows them to have negative energy states outside the ergosphere, a fact first noticed for the Reissner-Nordstr\"om BH leading to the concept of \textit{generalized ergosphere} \cite{1973PhLB...45..259D}. This property has been used in the extension of the Penrose process in the presence of magnetic fields (see, e.g., \cite{1984PhRvD..30.1625D,1986ApJ...307...38P,2019Univ....5..125T,2021PhRvD.103b4021K,2021PhRvD.104h4099T,2021Univ....7..416S}, and references therein).

However, the presence of electromagnetic fields does not guarantee an improvement in efficiency. An astrophysically relevant example is a Kerr BH immersed in a test magnetic field, asymptotically inclined relative to the BH rotation axis. Two seminal papers \cite{1976RSPSA.350..239P,1977RSPSA.356..351P} showed, in the slow-rotation regime and the full Kerr metric, that the exerted torque by the fields onto the BH induces the alignment between the BH angular momentum and the magnetic moment. The process fully converts the BH rotational energy into irreducible mass, with no energy being released to infinity at any time in the alignment process. The above result was also independently obtained using a different theoretical framework for the general Kerr BH metric in \cite{1977A&A....58..175K}.

In the meantime, several energy extraction mechanisms using electromagnetic fields have been proposed. The matter-dominated plasma accreting onto a Kerr BH \cite{1975PhRvD..12.2959R} inspired the Blandford-Znajek mechanism \cite{1977MNRAS.179..433B}, which has seen a recently boosted interest from relativistic magnetohydrodynamics and particle-in-cell simulations (e.g., \cite{2004MNRAS.350..427K,2005MNRAS.359..801K,2019PhRvL.122c5101P}). Further works involve magnetohydrodynamic inflows \cite{1990ApJ...363..206T}, electric and magnetic generalizations of the Penrose process \cite{1984PhRvD..30.1625D,1986ApJ...307...38P,2019Univ....5..125T,2021PhRvD.103b4021K,2021PhRvD.104h4099T,2021Univ....7..416S}, and references therein), and more recently, relativistic magnetic reconnection in the BH vicinity \cite{2021PhRvD.103b3014C}. Because of the critical role of the BH irreducible mass increase discussed above, it remains to verify whether or not the efficiency condition (\ref{eq:dMirrcond}) is approached in these processes. However, such a study goes beyond the scope of the present work and is left as a prospect. 

Next, we follow our plan of setting the physical picture of present interest to assess the conditions for the energy extraction to occur and its efficiency by monitoring the increase of the BH irreducible mass. 

\section{Electromagnetic field structure} \label{sec:3}

In spheroidal Boyer-Lindquist coordinates $(t, r, \theta, \phi)$, the Kerr metric reads \cite{1968PhRv..174.1559C} 
{
\begin{subequations}\label{eq:metric}
\begin{align}
    ds^2 &= g_{00} dt^2 + g_{11} dr^2 + g_{22} d\theta^2 + g_{33} d\phi^2 + 2 g_{0 3} dt d\phi,\\
    g_{00} &= -\left(1-\frac{2 M r}{\Sigma}\right),\quad g_{11} = \frac{\Sigma}{\Delta},\quad g_{22} = \Sigma,\\
    g_{33} &= \frac{A}{\Sigma}\sin^2\theta,\quad g_{0 3} = - \frac{2 a M r}{\Sigma} \sin^2\theta,
\end{align}
\end{subequations}
}
where , being $M$ and $a=J/M$, respectively, the BH mass and angular momentum per unit mass.

The electromagnetic four-potential of the Wald solution in the case of a slightly charged Kerr BH, embedded in a magnetic field of strength $B_0$ asymptotically aligned with the BH rotation axis is given by \cite{1974PhRvD..10.1680W},
\begin{equation}\label{eq:Amu}
    A_\mu =\frac{B_0}{2}\,\psi_\mu + a \,B_0 \eta_\mu - \frac{Q}{2 M}\eta_\mu,
\end{equation}
where $\eta^\mu= \delta^\mu_{\hphantom{\mu}{0}}$ and $\psi^\mu=\delta^\mu_{\hphantom{\mu}{3}}$ are the time-like and space-like Killing vectors of the Kerr metric. {Thus, the electromagnetic four-potential is $A_\mu = (A_0,0,0,A_3)$, where the non-vanishing components are given by (see \ref{app:A})}
\begin{subequations}\label{eq:potential}
    \begin{align}
    A_0 &= -a B_0 \Bigg[ 1 - \frac{M r}{\Sigma} (1+\cos^2\theta) \Bigg] +\frac{Q}{2 M}\left(1-\frac{2 M r}{\Sigma}\right),\\
    A_3 &= \frac{1}{2} B_0\sin^2\theta  \Bigg[ r^2 + a^2 - \frac{2 M r a^2}{\Sigma} (1+\cos^2\theta) \Bigg] \nonumber \\
    &+\frac{Q a r \sin^2\theta}{\Sigma}.
\end{align}
\end{subequations}

With the knowledge of $A_\mu$, we can now calculate the Faraday tensor, $F_{\alpha \beta} = \partial_\alpha A_\beta - \partial_\beta A_\alpha$, whose non-vanishing components result to be {(see \ref{app:A})}
\begin{subequations}\label{eq:Fab}
    \begin{align}
    F_{01} &= \frac{(r^2-a^2 \cos^2\theta)}{\Sigma^2}\left[a B_0 M (1+\cos^2\theta)-Q\right],\\
    F_{02} &= \frac{2 a r \sin\theta\cos\theta}{\Sigma^2}\left[ B_0 M (r^2-a^2)+ a Q\right]\\
    F_{13} &=B_0 r \sin^2\theta \left[ 1+ \frac{M a^2 (r^2-a^2 \cos^2\theta)(1+\cos^2\theta)}{r \Sigma^2}\right] \nonumber \\
    &-\frac{Q a (r^2 - a^2 \cos^2\theta)\sin^2\theta}{\Sigma^2},\\
    F_{23} &= \frac{B_0 \sin\theta\cos\theta}{\Sigma^2} \left[ \Sigma^2 (r^2 + a^2) -2 M a^2 r \Sigma (1+\cos^2\theta) \right.\nonumber \\
    & \left.+ 2 M a^2 r (r^2-a^2)\sin^2\theta \right] + \frac{2 Q a r^3 \sin\theta \cos\theta}{\Sigma^2}.
\end{align}
\end{subequations}

To better depict the electromagnetic field structure and the expected charged particle motion, we calculate the electromagnetic invariant {(see \ref{app:A})}
\begin{equation}\label{eq:EdotB}
    \vec{E}\cdot \vec{B} = -\frac{1}{4} F_{\alpha \beta} \tilde{F}^{\alpha\beta} = \frac{F_{02}F_{13}-F_{01}F_{23}}{\Sigma \sin\theta}, 
\end{equation}
where $\tilde{F}^{\alpha\beta}$ is the dual of the electromagnetic tensor, defined by $\sqrt{-g}\tilde{F}^{\alpha\beta} = (1/2)\epsilon^{\alpha \beta \mu \nu} F_{\mu \nu}$, being $\epsilon^{\alpha \beta \mu \nu}$ the Levi-Civita symbol, and $g = -\Sigma^2 \sin^2\theta$ the determinant of the Kerr spacetime metric (\ref{eq:metric}). 

\begin{figure*}
    \centering
    \includegraphics[width=\hsize,clip]{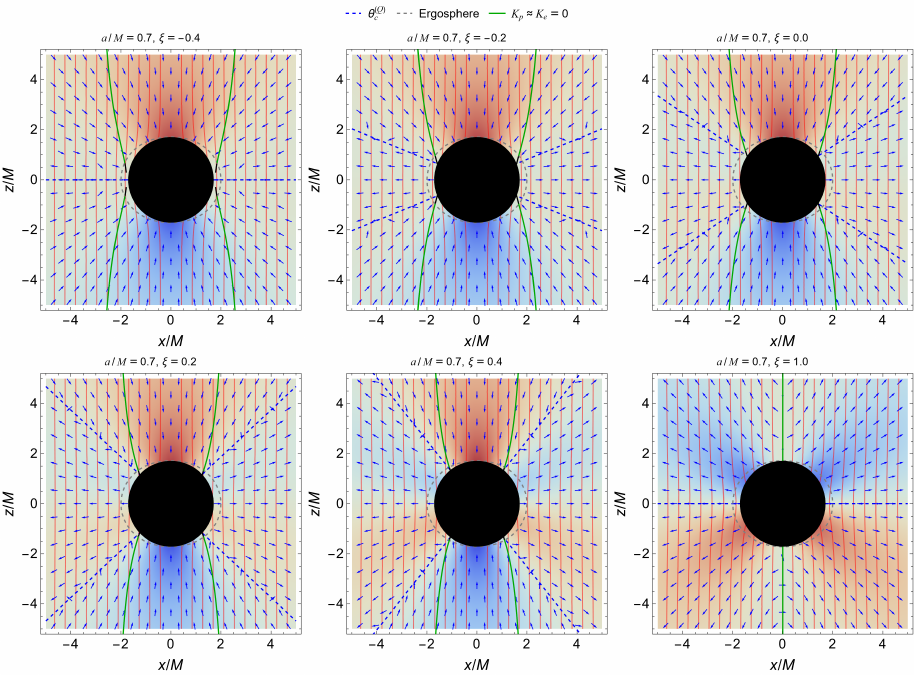}
    \caption{BH horizon (filled-black), ergosphere (dashed-gray), $K_p = 0$ boundary (green), electric field lines (blue arrows) and magnetic field lines (red, contours of constant $A_3$) for selected values of the charge parameter, $\xi = Q/Q_W$, where $Q_W = 2 J B_0 = 2 a M B_0$. By normalizing the radial coordinate to $M$ and the electric and magnetic field intensity to $B_0$, the EM field expressions depend only on $a/M$ and $\xi$. In this example, the BH spin parameter is $a/M = 0.7$, in the upper panel, $\xi = -0.4$ (left), $-0.2$ (center), and $0.0$ (right), and in the lower panel, $\xi = 0.2$ (left), $0.4$ (center), and $1.0$ (right). The lower panel shows how the angle at which $\vec{E}\cdot\Vec{B}=0$ (besides the equator), marked by the dashed-blue lines, shrinks with the increase of the positive charge, for instance, $\theta^{(Q)}_c\approx 56.12^\circ$ ($\xi=0$), $46.87^\circ$ ($\xi=0.2$), and $38.44^\circ$ ($\xi=0.4$). The background color maps the value of $\vec{E}\cdot\Vec{B}$, i.e., the redder color is negative and the bluer positive.}
    \label{fig:Kpzero}
\end{figure*}

{We are interested in obtaining the polar angle at which $\vec{E}\cdot \vec{B} = 0$, which} separates regions where the scalar product is positive and negative and can be calculated at the BH horizon radius. {By replacing the corresponding expressions of $F_{\alpha \beta}$ given in Eq. (\ref{eq:Fab}) into Eq. (\ref{eq:EdotB}), it follows} the quadratic equation $(\vec{E}\cdot \vec{B})_+ = \lambda Q^2 + \sigma Q + \gamma = 0$, whose two roots are
\begin{equation}\label{eq:thetaQ}
    Q = \frac{-\sigma \pm \sqrt{\sigma^2 - 4 \lambda \gamma}}{2 \lambda},
\end{equation}
where we have defined 
\begin{align}
    \lambda &= \frac{2 a r_+ (r_+^2-a^2 \cos^2\theta) \cos\theta}{\Sigma^4},\\
    \sigma &=-\frac{2 B_0 a^2 r_+ \cos\theta}{ \Sigma^4} [M (r_+^2-a^2\cos^2\theta)(1+\cos^2\theta) \nonumber \\
    &-r_+\sin^2\theta(r_+^2+a^2\cos^2\theta)],\\
    \gamma &= \sigma \frac{B_0 M (r_+^2-a^2)}{a}.
\end{align}

The background color in Fig. \ref{fig:Kpzero} shows the regions where the invariant (\ref{eq:EdotB}) is positive (bluer), negative (redder), and zero (blue dashed lines). For example, for a magnetic field aligned with the BH spin (i.e., vertically upward), positively charged particles will follow the magnetic field lines downward in the red and upward in the blue regions. Negatively charged particles have the opposite behavior. 

Given the invariant character of $\vec{E}\cdot \vec{B}$, we can use any observer to exemplify the above situation, e.g., the \textit{locally non-rotating} observer, also called the \textit{zero angular momentum} (ZAMO) observer \cite{1970ApJ...162...71B,1972ApJ...178..347B}. The ZAMO carries a tetrad with vectors $e_{\hat{0}}  = u_{(Z)}$, $e_{\hat{1}} =  \sqrt{\Delta/\Sigma}\, e_1$, $e_{\hat{2}} =e_2/\sqrt{\Sigma}$, and $e_{\hat{3}} =  \sqrt{\Sigma/A}\,e_3/\sin\theta$, where $u^\nu_{(Z)}$ the ZAMO four-velocity as seen for an observer at rest at infinity, $u^\nu_{(Z)} = \Gamma (1,0,0,\omega)$, with $\Gamma = \sqrt{A/(\Sigma \Delta)}$ and $\omega = 2 M a r/A$.

The electric and magnetic field components measured by the ZAMO are $E_{\hat{i}} = E_\mu\,e^\mu_{\hphantom{\mu}{\hat i}}$ and $B_{\hat{i}} = B_\mu\,e^\mu_{\hphantom{\mu}{\hat i}}$, where $E_\mu = F_{\mu \nu} u^\nu_{(Z)}$ and $B_\mu = \tilde{F}_{\mu \nu} u^\nu_{(Z)}$. Using the above ZAMO tetrad and the electromagnetic tensor components (\ref{eq:Fab}), the resulting electric and magnetic fields measured by the ZAMO {are $E_{\hat i} = (E_{\hat{1}},E_{\hat{2}},0)$ and $B_{\hat i} = (B_{\hat{1}},B_{\hat{2}},0)$, where}
\begin{subequations}\label{eq:Efieldzamo}
\begin{align}
E_{\hat{1}} &= -\frac{B_0 a M}{\Sigma^2 \sqrt{A}} \Bigg[(r^2 + a^2)(r^2-a^2\cos^2\theta)(1+\cos^2\theta) \nonumber \\
&- 2 r^2 \sin^2\theta\,\Sigma\Bigg] + Q\frac{(r^2 + a^2)(r^2 - a^2 \cos^2\theta)}{\Sigma^2 \sqrt{A}},\\
E_{\hat{2}} &= \frac{2 r a^2 \sin\theta \cos\theta}{\Sigma^2}\left[B_0 a M (1+\cos^2\theta)-Q\sqrt{\frac{\Delta}{A}}\right],
\end{align}
\end{subequations}
and
\begin{subequations}\label{eq:Bfieldzamo}
\begin{align}
B_{\hat{1}} &= -\frac{B_0 \cos\theta}{\Sigma^2 \sqrt{A}} \Bigg\{2 M r a^2 [2 r^2 \cos^2\theta+a^2(1+\cos^4\theta)] \nonumber \\
&-(r^2+a^2)\Sigma^2\Bigg\}+Q a\frac{2 r (r^2+a^2) \cos\theta}{\sqrt{A} \Sigma^2},\\
B_{\hat{2}}&= \sqrt{\frac{\Delta}{A}}\frac{\sin\theta}{\Sigma^2} \{-B_0 [M a^2 (r^2-a^2\cos^2\theta)(1+\cos^2\theta) \nonumber \\ &+ r \Sigma^2] + Q a (r^2 - a^2 \cos^2\theta)\}.
\end{align}
\end{subequations}
Figure \ref{fig:Kpzero} shows the electric and magnetic field lines in the ZAMO frame. In the northern and southern regions (hereafter polar regions), comprised at polar angles $-\theta^{(Q)}_c<\theta<\theta^{(Q)}_c$ and $\pi-\theta^{(Q)}_c<\theta<\pi+\theta^{(Q)}_c$, respectively, the electric field accelerates electrons outward and protons inward. In the eastern and western regions (hereafter equatorial regions), $\theta^{(Q)}_c\leq \theta \leq \pi-\theta^{(Q)}_c$ and $-\pi + \theta^{(Q)}_c\leq \theta \leq -\theta^{(Q)}_c$, respectively, the electric field accelerates electrons inward and protons outward. Notice that this result is independent of the ZAMO observer since it arises from the analysis of the Maxwell invariant associated with $\vec{E}\cdot \vec{B}$ (see Eq. \ref{eq:EdotB}).

Introducing the charge parameter, $\xi = Q/Q_W$, where $Q_W = Q_{\rm eff} = 2 J B_0 = 2 a M B_0$ is the so-called Wald charge (or effective charge; see \cite{2019ApJ...886...82R,2021A&A...649A..75M,2020EPJC...80..300R}, for this concept), and normalizing the radial coordinate to $M$ and the electric and magnetic field to $B_0$, by fixing a value of $a/M$, the angle $\theta^{(Q)}_c$ depends only on $\xi$. Figure \ref{fig:Kpzero} shows the electric field lines (blue arrows) and the magnetic field lines (contours of constant $A_\phi$, in red color) for a BH with $a/M = 0.7$. In this case, the solution of Eq. (\ref{eq:thetaQ}) gives, besides the trivial solution $\theta^{(Q)}_c = 90^\circ$, $\theta^{(Q)}_c\approx 56.12^\circ$ ($\xi=0$), $46.87^\circ$ ($\xi=0.2$), and $38.44^\circ$ ($\xi=0.4$), so the polar region shrinks with the increase of the positive charge. On the contrary, the increase of a negative charge enlarges it, e.g., $\theta^{(Q)}_c\approx 67.86^\circ$ ($\xi=-0.2$).

\section{Particle energy and angular momentum}\label{sec:4}

The conserved energy and angular momentum of massive, charged, test particle of mass $m_i$ and charge $q_i$ are 
\begin{subequations}\label{eq:EiLi}
\begin{align}
    E_i &= - \pi_\mu \eta^\mu = -\pi_0 = -p_0 - q A_0,\label{eq:Ei}\\
    L_i &= \pi_\mu \psi^\mu = \pi_3 = p_3 + q A_3,\label{eq:Li}
\end{align}
\end{subequations}
where $p_\alpha = m_i u_\alpha$ the four-momentum, $u_\alpha$ the four-velocity, $A_\mu$ is the electromagnetic four-potential given by Eqs. (\ref{eq:Amu}), and $i = p,e$ stands for protons or electrons.  We also refer the reader to the discussions of charged particle motion, energy, and angular momentum in the case of the Wald solution presented in \cite{2018PhRvD..98l3002L,2021PhRvD.104h4059G,2022MNRAS.512.2798K}. 

As in Ref. \cite{2023EPJC...83..960R}, we study test particles initially at rest at the position $(r_i, \theta_i, \phi_i)$, outside the ergosphere. Hence, $\Sigma_{i} > 2 M r_{i}$, with $r_{i} > r_{\rm erg} = M + \sqrt{M^2-a^2 \cos^2\theta_{i}}$, and the initial four-velocity is $u^\alpha_{i} = u^0_{i} \delta^\alpha_0$, with $u^0_{i} = (1-2 M r_{i}/\Sigma_{i})^{-1/2}$. The kinetic energy of a particle crossing the BH horizon is 
\begin{equation}\label{eq:Ki}
    K_i = -p_\mu l^\mu|_+ = E_i -\Omega_+ L_i.
\end{equation}
where $l^\mu = \eta^\mu + \Omega_+ \psi^\mu$ \cite{1973blho.conf...57C}. By inspecting Eq. (\ref{eq:dMirr}), it can be seen that the condition $K_i>0$ implies the increase of the square of the BH irreducible mass.

\begin{figure*}
    \centering
    \includegraphics[width=0.95\hsize,clip]{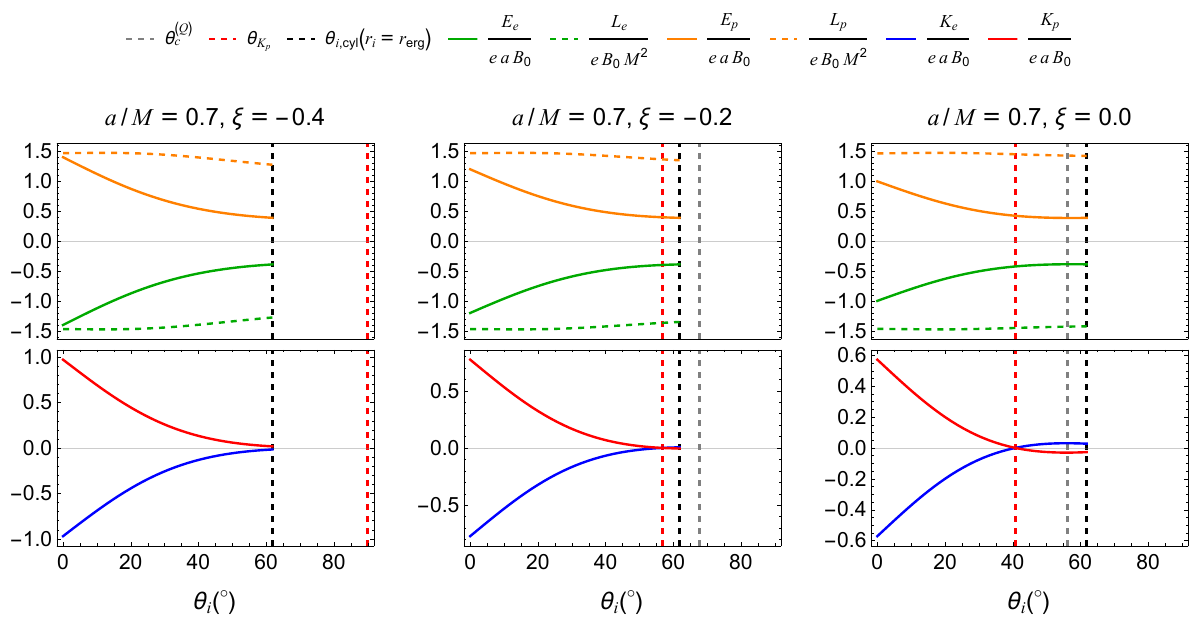}
    \includegraphics[width=0.95\hsize,clip]{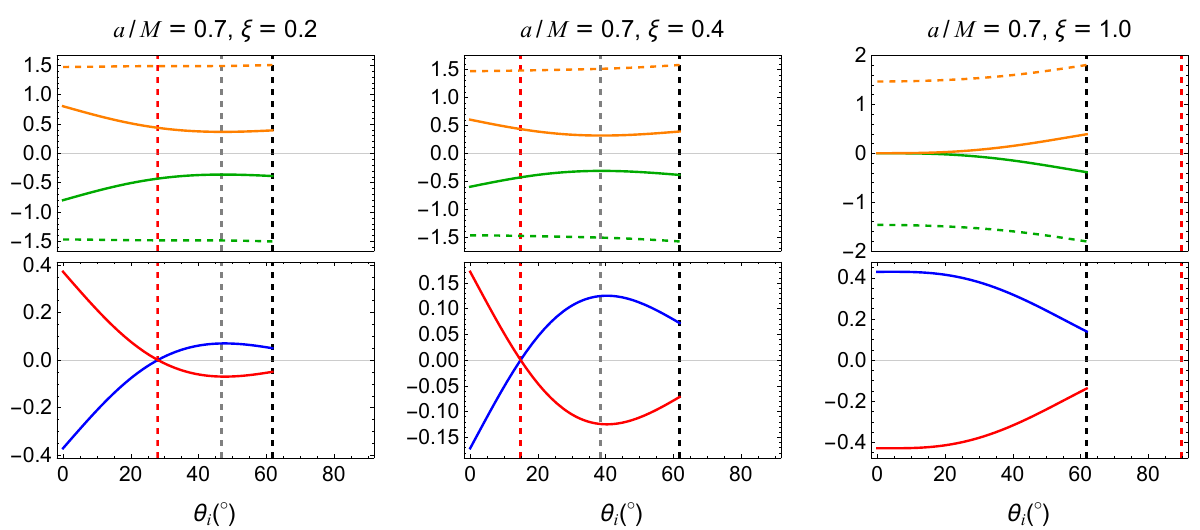}
    \caption{$E_i$, $L_i$, given by Eqs. (\ref{eq:Ei})--(\ref{eq:Li}), and $K_i$ given by Eq. (\ref{eq:Ki}), at initial positions outside the ergosphere ($r_i > r_{\rm erg}$) leading to the particle capture by the BH (see Fig. \ref{fig:Kpzero}). In this example, the initial radial coordinate is set by $r_i = r_H/\sin\theta_i$, and the polar angle by $0\leq\theta_i\leq \theta_{i,\rm cyl}(r_i = r_{\rm erg})$, and the polar angle is $\theta_{i,\rm cyl}(r_i = r_{\rm erg})\approx 61.86^\circ$ is the angle at which $r_i = r_{\rm erg}$.}
    \label{fig:EmOmL}
\end{figure*}

Charged particles will follow the magnetic field lines, which are nearly vertical (see Fig. \ref{fig:Kpzero}), so the BH can capture particles at initial positions $r_i\sin\theta_i \lesssim r_H$. For example, let us set initial particle positions $r_i\sin\theta_i = r_H$. The upper panel of Fig. \ref{fig:EmOmL} shows $E_e$, $L_e$, $E_p$, $L_p$, $K_e$, $K_p$, and indicates the spherical polar angles $\theta_c^{(Q)}$, $\theta_{K_p}$, and $\theta_{i, \rm cyl}(r_i = r_{\rm erg})$. The angle $\theta_{i, \rm cyl}(r_i = r_{\rm erg})$ is the maximum value at which the initial position is along the set cylinder and above the ergosphere, i.e., the solution of the equation $r_{\rm erg} \sin\theta_i = r_H$. The angle $\theta_{K_p}$ is that where $K_p = 0$. Thus, the BH captures polar protons (in the northern hemisphere) in the region $0\leq \theta \leq {\rm Min}(\theta_c^{(Q)}, \theta_{i, \rm cyl}, \theta_{K_p})$. Let us analyze the neutral case, $\xi = 0$ (upper right figure). We have $\theta_{K_p} \approx 41^\circ$, $\theta_c^{(Q)} \approx 57^\circ$, and $\theta_{i,\rm cyl}(r_i = r_{\rm erg})\approx 62^\circ$. The BH captures protons in the region $0\leq \theta_{K_p}$. These protons have positive energy and angular momentum (see solid and dashed orange curves). For larger angles, proton trajectories do not cross the event horizon (the red curve, $K_p$, becomes negative), and the BH captures electrons (the blue curve, $K_e$, becomes positive). Those electrons have negative energy and angular momentum (solid and dashed green curves). An analogous analysis can be done for non-zero values of the charge parameter: it turns out that the proton(electron) capture region shrinks(enlarges) for positive values of the charge parameter and vice-versa for negative values, as expected from Fig. \ref{fig:Kpzero}.

\section{BH mass, angular momentum, and irreducible mass change} \label{sec:5}

By capturing a particle, the BH mass and angular momentum change by {$d M_i = E_i$ and $d J_i = L_i$}. Because $d M_i - \Omega_+ d J_i = E_i - \Omega_+ L_i = K_i \geq 0$ (see Fig. \ref{fig:EmOmL}), we have $d M^2_{\rm irr} \geq 0$ (see Eq. \ref{eq:dMirr}), as expected from the BH reversible and irreversible transformations \cite{1970PhRvL..25.1596C,1971PhRvD...4.3552C} or the horizon surface area increase theorem \cite{1971PhRvL..26.1344H}.

For a particle number density $n$, we can estimate the net energy and angular momentum transferred {to the BH} by the captured particles as \cite{2023EPJC...83..960R}
{
\begin{subequations}\label{eq:ELtotal}
    \begin{align}
        {\cal E}_i = 2 \pi \iint E_i n \sqrt{-g}\,dr_i d\theta_i = 2 \pi \iint E_i n \Sigma_i \sin\theta_i\,dr_i d\theta_i , \label{eq:Eitotal}\\
        {\cal L}_i = 2 \pi \iint L_i n \sqrt{-g}\,dr_i d\theta_i =2 \pi \iint L_i n \Sigma_i \sin\theta_i\,dr_i d\theta_i\label{eq:Litotal},
    \end{align}
\end{subequations}
}
where the integration is carried out in the region {covering all initial positions} of particles captured by the BH, as discussed in the previous section. {We calculate the conserved energy and angular momentum at the initial position $(r_i,\theta_i)$ of the electrons ($i=e$) and protons ($i=p$), where we assume they start their motion at rest, so $u^0_e = u^0_p = 1/\sqrt{-g_{00}} = (1-2 M r_{e,p}/\Sigma_{e,p})^{-1/2}$. Hence, Eqs. (\ref{eq:EiLi}) become
\begin{subequations}\label{eq:EiLiexplicit}
    \begin{align}
    E_e &= m_e \sqrt{-g_{00}} + e A_0,\quad E_p = m_p \sqrt{-g_{00}} - e A_0,\\
    L_e &= m_e \frac{g_{03}}{\sqrt{-g_{00}}} - e A_3,\quad L_p = m_p \frac{g_{03}}{\sqrt{-g_{00}}} + e A_3,
\end{align}
\end{subequations}
where we have denoted with $m_{e,p}$ the electron and proton rest-mass and $q_e = -e$ and $q_p = +e$ their charge, and the electromagnetic four-potential components are given by Eqs. (\ref{eq:potential}). Therefore, we obtain the total energy and angular momentum absorbed by the BH are
\begin{equation}\label{eq:ELtotal2}
    {\cal E} = {\cal E}_e + {\cal E}_p,\quad {\cal L} = {\cal L}_e + {\cal L}_p
\end{equation}
where
\begin{subequations}\label{eq:ELtotalexplicit}
    \begin{align}
        {\cal E}_e &= 2 \pi \iint n \Sigma_e \sin\theta_e \left[ m_e \sqrt{-g_{00}} + \frac{e B_0}{2}g_{03} \right.\nonumber \\
        &+ \left. \left( e a B_0 - \frac{e Q}{2 M}\right)g_{00}\right]\,dr_ e\theta_e , \label{eq:Eetotal}\\
        {\cal L}_e &= 2 \pi \iint n \Sigma_e \sin\theta_e \left[m_e \frac{g_{03}}{\sqrt{-g_{00}}} - \frac{e B_0}{2}g_{33} \right.\nonumber \\
        &-\left. \left( e a B_0 - \frac{e Q}{2 M}\right)g_{03}\right]\,dr_ e\theta_e , \label{eq:Letotal}\\
        {\cal E}_p &= 2 \pi \iint n \Sigma_p \sin\theta_p \left[ m_p \sqrt{-g_{00}} - \frac{e B_0}{2}g_{03} \right.\nonumber \\
        &- \left.\left( e a B_0 - \frac{e Q}{2 M}\right)g_{00}\right]\,dr_ p\theta_p , \label{eq:Eptotal}\\
        {\cal L}_p &= 2 \pi \iint n \Sigma_p \sin\theta_p \left[m_p \frac{g_{03}}{\sqrt{-g_{00}}} \right.\nonumber \\
        &+ \left.\frac{e B_0}{2}g_{33} + \left( e a B_0 - \frac{e Q}{2 M}\right)g_{03}\right]\,dr_ p\theta_p,\label{eq:Lptotal}
    \end{align}
\end{subequations}
where we have used the expression of $A_\mu$ given in Eq. (\ref{eq:Amuexplicit}) in \ref{app:A}, Eqs. (\ref{eq:EiLiexplicit}), and the involved metric functions are given in Eqs. (\ref{eq:metric}).
}

{With the above, the total change of the mass, angular momentum, and irreducible mass are
\begin{align}\label{eq:DeltaML}
    \Delta M &= {\cal E},\quad \Delta J = {\cal L}, \quad \Delta M_{\rm irr} = M_{\rm irr} \frac{{\cal E}-\Omega_+ {\cal L}}{\sqrt{M^2-a^2}},
\end{align}
where we have assumed the changes as infinitesimal (relative to the BH initial parameters) and used Eq. (\ref{eq:Mirr}) to estimate the change of the irreducible mass. 
}

\begin{figure*}
    \centering
    \includegraphics[width=\hsize,clip]{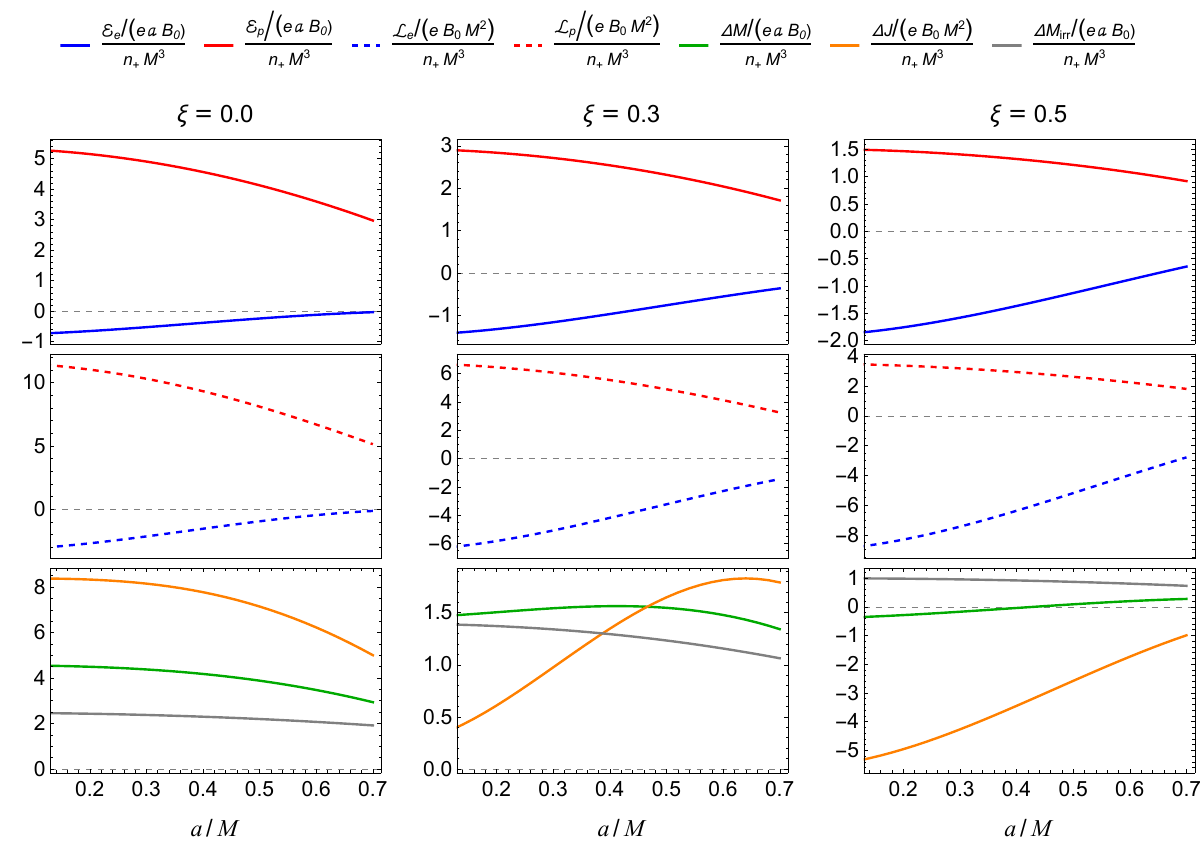}
    \caption{{Upper and middle panels: total energy and angular momentum, ${\cal E}_i/(e B_0 a)$ and ${\cal L}_i/(e B_0 M^2)$, of the polar protons (solid and dashed red) and equatorial electrons (solid and dashed blue) absorbed by the BH, given by Eqs. (\ref{eq:ELtotalexplicit}). Lower panels: The net change of the BH mass-energy $\Delta M$ (green), angular momentum $\Delta L$ (orange), and irreducible mass $\Delta M_{\rm irr}$ (gray), according to Eq. (\ref{eq:DeltaML})}.}
    \label{fig:gridtotal}
\end{figure*}

Figure \ref{fig:gridtotal} shows {in the upper panels ${\cal E}_e$ and ${\cal E}_p$, in the middle panels, ${\cal L}_e$ and ${\cal L}_p$, and in the lower panels, $\Delta M$, $\Delta J$, and $\Delta M_{\rm irr}$, for selected values of the charge parameter, $\xi = 0$ (first column), $0.3$ (central column), $0.5$ (last column)}, as a function of the BH spin parameter. We assume a spherically symmetric density, $n(r) = n_+ (r_+/r)^m$, with $m = 3$. {It is clear from Eq. (\ref{eq:ELtotalexplicit}) that the parameter $n_+$ is only a scaling factor, so we do not need to specify it for the general discussion. However, we must remember that the particle density around the BH must be lower than the Goldreich-Julian density to avoid the screening of the accelerating electric field (see discussion in \cite{2023EPJC...83..960R}). For instance, for a BH of mass $M=4 M_\odot$, spin $a = M$, surrounded by a magnetic field of $10^{13}$ G, the density must be lower than a few $10^{-9}$ g cm$^{-3}$, so $n_+ \lesssim 10^{15}$ cm$^{-3}$. Further, for these astrophysical BH and magnetic field parameters, the electric potential energy largely dominates over the gravitational one \cite{2023EPJC...83..960R}, which implies, from Eq. (\ref{eq:EiLiexplicit}), that $E_{e,p} \approx \pm e A_0$ and $L_{e,p} \approx \mp e A_3$, which do not depend upon the particle mass. Under these conditions, and recalling that $Q = \xi Q_W = \xi 2 a M B_0$, $E_{e,p}$ and $L_{e,p}$ scale with $e a B_0$ and $e B_0 M^2$, respectively, which explains the normalization used in Figure \ref{fig:gridtotal}.} 

The figure suggests the BH could follow a cyclic behavior. Let us start with a neutral BH ($\xi = 0$, left column plots). {As already discussed in \cite{2023EPJC...83..960R}, and shown by this plot, the BH in this initial phase absorbs more protons than electrons, so it gains energy and angular momentum ($\Delta M > 0$ and $\Delta J > 0$), and becomes positively charged ($\xi >0$). Figures \ref{fig:Kpzero} and \ref{fig:EmOmL} show that, for $\xi >0$, the polar region shrinks (the equatorial enlarges), so the positive contribution of protons to the energy and angular momentum reduces, and the negative one of electrons increases, as shown by the central column panels. The process can continue this way until the BH starts to absorb more electrons than protons, and it gains net negative energy and angular momentum ($\Delta M < 0$ and $\Delta J < 0$), as shown in the last column panels. In this stage, the BH energy is extracted. It can be shown that for $\xi > \xi_c = (3-\sqrt{1-(a/M)^2})/2$, the polar region is entirely contained in the cylinder of radius $r_+$, leading to a considerable reduction of absorbed protons. The electrons can take over, causing the BH charge to reduce. The BH energy extraction continues. However, the decrease in the BH charge reverses the shrinking of the polar region and the equatorial region enlargement. The charge becomes again $\xi<\xi_c$, and at some instant, the enlargement of the polar region is such that protons take over once again. The BH energy extraction stops, and the cycle repeats.}

\section{Conclusions}\label{sec:6}

{We have generalized the analysis of Ref. \cite{2023EPJC...83..960R} of the energy extraction from a Kerr BH, immersed in a magnetic field asymptotically aligned to the BH spin, which captures positively and negatively charged particles from its surroundings, considering the effects of the BH charge. For this task, we used the charged Wald solution \cite{1974PhRvD..10.1680W}, i.e., for test BH charge. We have estimated the change of the BH mass, angular momentum, and irreducible mass for different values of the BH spin and charge parameter.}

{We have shown the changes of the region of capturable protons and electrons as a function of the BH charge parameter (see Figs. \ref{fig:Kpzero} and \ref{fig:EmOmL}). The changing polar and equatorial regions with the charge parameter leads to new results relative to the previously uncharged analysis in \cite{2023EPJC...83..960R}.} 

{For the present case, energy extraction occurs for a particle density that is isotropic and falls with distance as a power-law, e.g., $n \propto 1/r^3$, unlike the uncharged case \cite{2023EPJC...83..960R}. The increase of the irreducible mass in this process} is relatively low compared to purely gravitational processes like the Penrose process recently evaluated in Paper II.

For isotropic ionized matter density, our analysis for various spin and charge values suggests {the BH could evolve in cycles where its charge increases and decreases, alternating periods} without and with energy and angular momentum extraction (see Figs. \ref{fig:EmOmL} and \ref{fig:gridtotal}).

As we have made some assumptions, {it is worth mentioning possible} extensions and generalizations of this work. Two main extensions are particularly promising and relevant. First, abandoning the assumption of the BH charge as a test charge. Such an extension would allow us to verify how much the maximum extractable energy of $50\%$ of the energy of an extremely charged BH is approachable. Second, in our estimate of the increase of the irreducible mass, we have carried out the integrals (\ref{eq:ELtotal}) over the entire region where the charged particles can be captured by the BH of some mass and angular momentum. The above method implicitly assumes all those particles are captured simultaneously. Thus, an improvement would be calculating the equal capture time regions. The above extension is necessary for understanding the dynamic behavior of this system. {Further, they will lead to an improved assessment of the BH irreducible mass increase} in time, hence the efficiency of the energy extraction process.

\appendix

\section{{Expressions of the electromagnetic four-potential, Faraday tensor and its dual}}\label{app:A}

{
This appendix shows the explicit, full expressions, i.e., vanishing and non-vanishing components, of the electromagnetic four-potential, $A_\mu$, the Faraday tensor, $F_{\alpha \beta}$, and its dual, $\tilde{F}_{\alpha \beta}$. We start with $A_\mu$, so for this task, following Eq. (\ref{eq:Amu}), we must know the covariant and contravariant time-like and space-like Killing vectors
\begin{subequations}\label{eq:killingvectors}
    \begin{align}
    \eta^\mu &= \delta^\mu_0, \quad \eta_\mu = g_{\mu \alpha} \eta^\alpha = g_{\mu 0} = g_{00} \delta^0_\mu + g_{0 3} \delta^3_\mu,\\
    \psi^\mu &= \delta^\mu_3, \quad \psi_\mu = g_{\mu \alpha} \psi^\alpha = g_{\mu 3} = g_{0 3} \delta^0_\mu + g_{3 3} \delta^3_\mu.
\end{align}
\end{subequations}
Using Eqs. (\ref{eq:killingvectors}), $A_\mu$ can be written as
\begin{align}\label{eq:Amuexplicit}
    A_\mu &= \left[\frac{B_0}{2} g_{03} + \left( a B_0 - \frac{Q}{2 M} \right) g_{00}\right] \delta^0_\mu \nonumber \\
    &+ \left[\frac{B_0}{2} g_{33} + \left( a B_0 - \frac{Q}{2 M} \right) g_{0 3}\right] \delta^3_\mu,
\end{align}
where the metric tensor is
\begin{equation}
    g_{\alpha \beta} = 
    \left( 
    \begin{array}{cccc}
      g_{00} & 0 & 0 & g_{0 3}  \\
        0    & g_{11} & 0 & 0 \\
        0    & 0 & g_{22} & 0\\
        g_{03} & 0 & 0 & g_{3 3}
    \end{array}
    \right)
\end{equation}
with the components given in Eq. (\ref{eq:metric}). The inverse metric tensor is
\begin{equation}\label{eq:inversemetricmatrix}
    g^{\alpha \beta} = 
    \left( 
    \begin{array}{cccc}
      g^{00} & 0 & 0 & g^{0 3}  \\
        0    & g^{11} & 0 & 0 \\
        0    & 0 & g^{22} & 0\\
        g^{03} & 0 & 0 & g^{3 3}
    \end{array}
    \right),
\end{equation}
where
\begin{subequations}\label{eq:inversemetriccomp}
    \begin{align}
        g^{00} &= \frac{g_{33}}{g_{00} g_{33}-g^2_{03}} = -\frac{A}{\Sigma \Delta},\\
        g^{11} &= \frac{1}{g_{11}} = \frac{\Delta}{\Sigma},\\
        g^{22} &= \frac{1}{g_{22}} = \frac{1}{\Sigma},\\
        g^{33} &= \frac{g_{00}}{g_{00} g_{33}-g^2_{03}}=\frac{\Delta-a^2 \sin^2\theta}{\Sigma \Delta \sin^2\theta},\\
        g^{03} &= -\frac{g_{03}}{g_{00} g_{33}-g^2_{03}} = -\frac{2 M a r}{\Sigma \Delta}.
    \end{align}
\end{subequations}
Thus, by replacing the metric functions in Eq. (\ref{eq:Amuexplicit}), we obtain the electromagnetic four-potential
\begin{equation}\label{eq:Amuexplicit2}
    A_\mu = (A_0,0,0,A_3),
\end{equation}
where $A_0$ and $A_3$ are given in Eq. (\ref{eq:potential}).}

{
The Faraday tensor is defined as $F_{\alpha \beta} = \partial_\alpha A_\beta - \partial_\beta A_\alpha$. Thus, using Eq. (\ref{eq:Amuexplicit2}), and that $A_0$ and $A_3$ are only functions of $r$ and $\theta$ because of the axial symmetry, we obtain
\begin{align}
    F_{\alpha \beta} &= \partial_1 A_0 (\delta^1_\alpha \delta^0_\beta - \delta^1_\beta \delta^0_\alpha) + \partial_2 A_0 (\delta^2_\alpha \delta^0_\beta - \delta^2_\beta \delta^0_\alpha) \nonumber \\
    &+ \partial_1 A_3 (\delta^1_\alpha \delta^3_\beta - \delta^1_\beta \delta^3_\alpha) + \partial_2 A_3 (\delta^2_\alpha \delta^3_\beta - \delta^2_\beta \delta^3_\alpha),
\end{align}
which can be written in matrix form as
\begin{equation}\label{eq:Fabmatrix}
    F_{\alpha \beta} = 
    \left( 
    \begin{array}{cccc}
      0 & F_{01} & F_{02} & 0  \\
        F_{10}    & 0 & 0 & F_{13} \\
        F_{20}    & 0 & 0 & F_{23}\\
        0 & F_{31} & F_{32} & 0
    \end{array}
    \right),
\end{equation}
where 
\begin{subequations}
    \begin{align}
        F_{01} &= - F_{10} = - \partial_1 A_0,\quad F_{02} = - F_{20} = - \partial_2 A_0,\\
         F_{13} &= - F_{31} = \partial_1 A_3,\quad F_{23} = - F_{32} =  \partial_2 A_3,
    \end{align}
\end{subequations}
leading to the expressions given by Eqs. (\ref{eq:Fab}). For completeness, we give the expressions of the two-contravariant Faraday tensor $F^{\alpha \beta} = g^{\alpha \mu} g^{\beta \nu} F_{\mu \nu}$, in matrix form
\begin{equation}\label{eq:Fabupmatrix}
    F^{\alpha \beta} = 
    \left( 
    \begin{array}{cccc}
      0 & F^{01} & F^{02} & 0  \\
        F^{10}     & 0 & 0 & F^{13}  \\
       F^{20}    & 0 & 0 & F^{23}\\
        0 & F^{31} & F^{32} & 0
    \end{array}
    \right),
\end{equation}
where 
\begin{subequations}
    \begin{align}
        F^{01} &= -F^{10} = g^{11}(g^{00}F_{01} + g^{03}F_{31}),\\
        F^{02} &= -F^{20} = g^{22}(g^{00}F_{02} + g^{03}F_{32}),\\
        F^{13} &= -F^{31} = g^{11}(g^{03}F_{10} + g^{33}F_{13}),\\
        F^{23} &= -F^{32} = g^{22}(g^{03}F_{20} + g^{33}F_{23}),\\
    \end{align}
\end{subequations}
with the inverse metric given by Eqs. (\ref{eq:inversemetricmatrix}) and (\ref{eq:inversemetriccomp}).
}

{
For the determination of the electromagnetic scalars, we must compute the electromagnetic dual tensor, defined by $\sqrt{-g}\tilde{F}^{\alpha\beta} = (1/2)\epsilon^{\alpha \beta \mu \nu} F_{\mu \nu}$, being $\epsilon^{\alpha \beta \mu \nu}$ the Levi-Civita symbol given by
\begin{equation}
    \epsilon^{\alpha \beta \mu \nu} =  \left\{\begin{array}{rl}
        +1, & \text{for even permutations of 0,1,2,3}\\
        -1, & \text{for odd permutations of 0,1,2,3}\\
         0, & \text{otherwise}.
    \end{array}
    \right.
\end{equation}
With the above definition, we obtain
\begin{equation}\label{eq:Fabdualmatrix}
    \sqrt{-g}\tilde{F}^{\alpha\beta} = 
    \left( 
    \begin{array}{cccc}
      0 & F_{23} & -F_{13} & 0  \\
     -F_{23}  & 0 & 0 & -F_{02} \\
     F_{13}    & 0 & 0 & F_{01}\\
        0 & F_{02} & -F_{01} & 0
    \end{array}
    \right).
\end{equation}
It is straightforward to check that the matrix product of Eqs. (\ref{eq:Fabmatrix}) and (\ref{eq:Fabdualmatrix}) leads to the electromagnetic invariant (\ref{eq:EdotB}).
}


\end{document}